\documentclass[]{spie}  

\usepackage{amsmath,amsfonts,amssymb}
\usepackage{graphicx}
\usepackage[colorlinks=true, allcolors=blue]{hyperref}
\usepackage[utf8]{inputenc}
\usepackage{float}
\usepackage[T1]{fontenc}
\usepackage{booktabs}

\title{Engineering of titanium transition edge sensor wafers for the BA4-90/150 receiver of BICEP Array}

\author[a, *]{A.~Patel}
\author[b]{P.~A.~R.~Ade}
\author[c,d]{Z.~Ahmed}
\author[e]{M.~Amiri}
\author[f]{D.~Barkats}
\author[a]{R.~Basu~Thakur}
\author[g]{C.~A.~Bischoff}
\author[h]{D.~Beck}
\author[a,i]{J.~J.~Bock}
\author[j]{V.~Buza}
\author[h,c]{B.~Cantrall}
\author[a]{J.~R.~Cheshire~IV}
\author[k]{J.~Connors}
\author[l]{J.~Cornelison}
\author[m]{M.~Crumrine}
\author[a]{A.~J.~Cukierman}
\author[k]{E.~Denison}
\author[n]{L.~Duband}
\author[f]{M.~A.~Echter}
\author[o]{M.~Eiben}
\author[f,p]{B.~D.~Elwood}
\author[a]{S.~Fatigoni}
\author[q]{J.~P.~Filippini}
\author[h]{A.~Fortes}
\author[a]{M.~Gao}
\author[g]{C.~Giannakopoulos}
\author[h]{N.~Goeckner-Wald}
\author[h]{D.~C.~Goldfinger}
\author[r,s]{S.~Gratton}
\author[h]{J.~A.~Grayson}
\author[a]{A.~Greathouse}
\author[f]{P.~K.~Grimes}
\author[e]{M.~Halpern}
\author[c,d]{S.~Henderson}
\author[m]{T.~D.~Hoang}
\author[k]{J.~Hubmayr}
\author[a]{H.~Hui}
\author[h]{K.~D.~Irwin}
\author[t]{M.~Izquierdo~Poza}
\author[a]{J.~H.~Kang}
\author[t]{K.~S.~Karkare}
\author[a]{S.~Kefeli}
\author[f,p]{J.~M.~Kovac}
\author[h]{C.~Kuo}
\author[m,u]{K.~Lasko}
\author[a]{K.~Lau}
\author[g]{M.~Lautzenhiser}
\author[h]{T.~Liu}
\author[j,v]{S.~C.~Mackey}
\author[m]{N.~Maher}
\author[i]{K.~G.~Megerian}
\author[a]{L.~Minutolo}
\author[a]{L.~Moncelsi}
\author[h]{Y.~Nakato}
\author[a,i]{H.~T.~Nguyen}
\author[a,i]{R.~O’Brient}
\author[f]{S.~N.~Paine}
\author[f]{M.~A.~Petroff}
\author[f,p]{A.~R.~Polish}
\author[n]{T.~Prouve}
\author[m]{C.~Pryke}
\author[k]{C.~D.~Reintsema}
\author[a]{T.~Romand}
\author[h]{M.~Salatino}
\author[a]{A.~Schillaci}
\author[f]{B.~Schmitt}
\author[m,u]{B.~Singari}
\author[a,i]{A.~Soliman}
\author[f]{T.~St.~Germaine}
\author[a]{A.~Steiger}
\author[a]{B.~Steinbach}
\author[b]{R.~Sudiwala}
\author[h,c]{K.~L.~Thompson}
\author[b]{C.~Tucker}
\author[i]{A.~D.~Turner}
\author[w]{C.~Verg\`{e}s}
\author[j,v]{A.~G.~Vieregg}
\author[a]{A.~Wandui}
\author[i]{A.~C.~Weber}
\author[m]{J.~Willmert}
\author[a,c,d]{W.~L.~K.~Wu}
\author[h]{H.~Yang}
\author[j,l]{C.~Yu}
\author[f]{L.~Zheng}
\author[c]{C.~Zhang}
\author[a]{S.~Zhang}

\affil[a]{Department of Physics, California Institute of Technology, 1200 E. California Boulevard, Pasadena, CA 91125, USA}
\affil[b]{School of Physics and Astronomy, Cardiff University, Cardiff, CF24 3AA, United Kingdom}
\affil[c]{Kavli Institute for Particle Astrophysics and Cosmology, Stanford University, Stanford, CA 94305, USA}
\affil[d]{SLAC National Accelerator Laboratory, Menlo Park, CA 94025, USA}
\affil[e]{Department of Physics and Astronomy, University of British Columbia, Vancouver, British Columbia, V6T 1Z1, Canada}
\affil[f]{Center for Astrophysics, Harvard \& Smithsonian, Cambridge, MA 02138, USA}
\affil[g]{Department of Physics, University of Cincinnati, Cincinnati, OH 45221, USA}
\affil[h]{Department of Physics, Stanford University, Stanford, California 94305, USA}
\affil[i]{Jet Propulsion Laboratory, California Institute of Technology, 4800 Oak Grove Drive, Pasadena, CA 91109, USA}
\affil[j]{Kavli Institute for Cosmological Physics, University of Chicago, 5640 S Ellis Ave, Chicago, IL 60637, USA}
\affil[k]{National Institute of Standards and Technology, Boulder, CO 80305, USA}
\affil[l]{High-Energy Physics Division, Argonne National Laboratory, 9700 South Cass Avenue., Lemont, IL, 60439, USA}
\affil[m]{School of Physics and Astronomy, University of Minnesota, Minneapolis, MN 55455, USA}
\affil[n]{Service des Basses Temperatures, Commissariat a l’Energie Atomique, 38054 Grenoble, France}
\affil[o]{School of Engineering and Natural Sciences, Sæmundargata 2, 102 Reykjavík, Iceland}
\affil[p]{Department of Physics, Harvard University, Cambridge, MA 02138, USA}
\affil[q]{Department of Physics, University of Illinois at Urbana-Champaign, Urbana, Illinois 61801, USA}
\affil[r]{Centre for Theoretical Cosmology, DAMTP, University of Cambridge, Wilberforce Road, Cambridge CB3 0WA, UK}
\affil[s]{Kavli Institute for Cosmology Cambridge, Madingley Road, Cambridge CB3 0HA, UK}
\affil[t]{Department of Physics, Boston University, 590 Commonwealth Avenue, Boston, MA 02215, USA}
\affil[u]{Minnesota Institute for Astrophysics, University of Minnesota, Minneapolis, MN 55455, USA}
\affil[v]{Department of Physics, University of Chicago, 5720 S Ellis Ave, Chicago, IL 60637, USA}
\affil[w]{Lawrence Berkeley National Laboratory, 1 Cyclotron Road, Berkeley, CA 94720, USA}

\authorinfo{Further author information: (Send correspondence to A.P)\\A.P. E-mail: aapatel@caltech.edu}

\begin{document} 
\maketitle

\begin{abstract}
BA4-90/150, the fourth receiver to be deployed in the BICEP Array (BA) series, is a dichroic 90/150 GHz instrument targeting the frequency space where sensitivity to the CMB polarization is maximized. The receiver will be deployed in the 2026–2027 austral summer, and is set to position BA to achieve exceptionally precise measurements of cosmic microwave background (CMB) polarization and strengthen constraints on inflationary models. Recent measurements in existing BA receivers suggest that unexpectedly high loop gain in the titanium (Ti) transition edge sensors (TESs) produces excess high-frequency noise that is consequently aliased down into the science band through the time-division multiplexed readout. To reduce the loop gain, we fabricated and tested prototype Ti TES wafers containing 16 modified detector architectures designed to broaden the superconducting transition and reduce the transition steepness ($\alpha$). We present detector performance results, which will directly inform the final integrated wafer now being designed for full receiver commissioning.
\end{abstract}


\section{INTRODUCTION}

\begin{figure}[!t]
\centering
\includegraphics[width=2.5in]{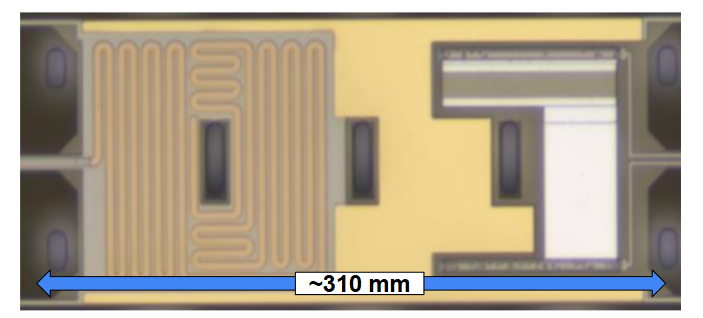}
\caption{Optical microscope image of a BA TES detector and meander. Credit: JPL}
\label{fig:BA TES SEM Image}
\end{figure}

The theory of cosmic inflation describes the period at approximately  $10^{-35}$ seconds after the Big Bang where there was a burst of exponential expansion. Proof of inflation would resolve the fine-tuned initial conditions in standard hot big bang cosmology. Inflation predicts the existence of primordial gravitational waves (PGW), which would imprint a characteristic ‘B-mode’ polarization in the CMB \cite{Kamionkowski2016,Seljak1997,SeljakZaldarriaga1997,Abazajian2016}. Detecting B-modes would provide the evidence required for inflation theory and offer scientists a rare window into grand unified theory scale physics present at the start of the Universe. The hunt for the primordial B-mode signal is one that spans decades, with experiments globally pursuing its detection. The BICEP experiments currently have the best constraint on r, and have ruled out inflation models that were previously favored, whilst simultaneously progressing our understanding of high energy physics \cite{Ade2021}.
BICEP Array (BA) is the latest series of receivers in the BICEP program. It consists of four small aperture receivers located at the South Pole, Antarctica. BA receivers collectively span frequency bands from $30~\mathrm{GHz}$ to $270~\mathrm{GHz}$. \cite{Hui2018,Dumarchez2024,KCollaboration2024}.
BA4-90/150 is the fourth receiver to be deployed in the BA series. As a dichroic 90/150 GHz instrument, it will target the frequency space where sensitivity to the CMB polarization is maximized. BA4-90/150 is central to achieving the program’s goal of $\sigma(r)$ on the order of 0.001 by 2034, in collaboration with the South Pole Telescope (SPT) for delensing \cite{Ade2021Delensing}.
At the core of each receiver sits the transition edge sensor (TES) detectors \cite{Irwin1995} responsible for sensing the incoming photons in order to detect the faint B-mode imprint of PGWs. An optical microscope image of a TES can be seen in Fig.~\ref{fig:BA TES SEM Image}. Detecting this faint signal requires exploitation of the detector transition region, which sits between the superconducting and normal states. Here, sensitivity is maximized and noise minimized.

Recent analysis of the detectors in the $150~\mathrm{GHz}$ (BA2-150) and $220~\mathrm{GHz}$ (BA3-220/270) receivers shows higher-than-expected noise equivalent temperatures (NET). The leading theory to explain this is that an unexpectedly high loop gain in the TES is causing excess high-frequency noise to be aliased down into the science band during the time-division multiplexed (TDM) readout \cite{Fatigoni2024}. Loop gain $\mathcal{L}$ quantifies the strength of the electrothermal feedback:
\begin{equation}
\label{eq:loopgain}
\qquad \mathcal{L} = \frac{\alpha P_{\text{electrical}}}{G T_c},
\end{equation}
where $P_{\text{electrical}}$ is the electrical power required to push the detectors into transition, $G$ is the thermal conductance to the bath, and $T_c$ is the superconducting transition temperature.

In this case, decreasing loop gain reduces the excess aliased noise and thus the NETs. This can be done by decreasing the $\alpha$ parameter or $P_{\text{electrical}}$, or a combination of the two. This paper discusses a way in which $\alpha$ was decreased by changing the TES geometry. $\alpha$ is defined as the logarithmic sensitivity of the TES resistance to temperature in the transition region:
\begin{equation}
\label{eq:alpha}
\alpha = \frac{T}{R}\frac{dR}{dT}
\end{equation}

As demonstrated in Eq.~\ref{eq:loopgain} and Eq.~\ref{eq:alpha}, a steep transition corresponds to a higher $\alpha$, producing stronger electrothermal feedback to destabilize the detector, thus leading to excess high-frequency noise, which gets aliased into the signal band. The BA TES’s have an $\alpha$ design target of $\sim$100, but recent measurements measure $\alpha$ at around a factor of five times greater. This motivated the design and fabrication of two test wafers with a combined sum of 16 different TES architectures that would investigate different recipes to soften the $\alpha$ transition, pushing it closer to its prescribed value. These detectors were intentionally left unreleased by skipping the final release etch step, which thermally isolates the TES island from the substrate. Because measuring $\alpha$ does not require released detectors, this approach shortens the fabrication process. The results will directly guide the fabrication of the remaining science-grade tiles for the BA3-220/270 receiver, and the upcoming BA4-90/150 receiver with the improved recipe for a softer $\alpha$. 

In parallel, we have experimented with decreasing $P_{\text{electrical}}$ through elevating our focal plane temperature. Increasing the bath temperature supplies additional power to the detectors, thereby reducing the $P_{\text{electrical}}$ required to push the detectors into transition at their operating bias.

\section{Design strategies to decrease $\alpha$}

Each architecture attempts to reduce $\alpha$ using the same physical lever of introducing spatial inhomogeneity into the transition, thus causing different regions of the TES film to pass through transition at very slightly different temperatures, and consequently broadening the transition.  Instead of the film transitioning from superconducting to normal as one at a single sharp point, the transition is spread across a wider temperature range. $dR/dT$ is lowered since the $R(T)$ curve is broadened, and so $\alpha$ is softened. This approach builds upon prior TES stability engineering efforts \cite{George2014}. 
The Microdevices Laboratory at the Jet Propulsion Laboratory fabricated two prototype wafers, which cumulatively held 16 different unreleased TES architectures and are referred to in this paper as $\alpha$1 through $\alpha$16. Through isolation and analysis of each architecture, we were able to characterize their $\alpha$ value, as well as other properties such as critical temperature ($T_c$), and normal resistance ($R_{\rm n}{\rm Ti}$). 
\subsection {Inverse Stack Ti-First Recipe Flavors}
\begin{figure}[!t]
\centering
\includegraphics[width=0.45\linewidth]{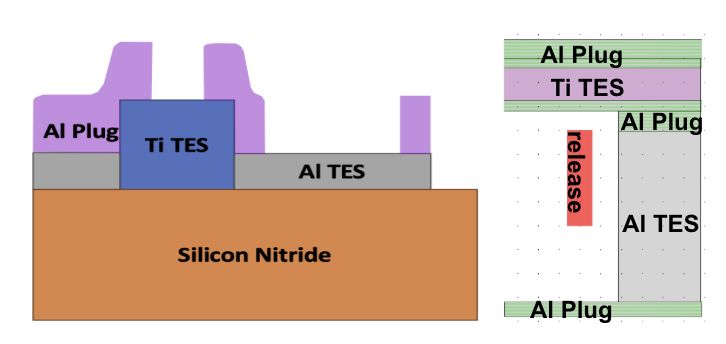}
\caption{(Left) Cross-sectional schematic of the $\alpha$1 TES stack showing the silicon nitride membrane (orange), Al calibration TES (gray), Ti science TES (blue), Al-plug (purple) layer. (Right) Top-view of an unreleased $\alpha$1 detector showing the Al-plug (green), Ti TES (purple), Al calibration TES (gray), and the release (red), which is removed during the final release etch. Credit: JPL, A. Turner}
\label{fig:BA Cartoon2}
\end{figure}

All architectures in Wafer 1 ($\alpha$1 through $\alpha$9) use the Ti-first recipe. This is the current fabrication stack for the BA detectors, as seen in the cartoon cross-section on the left of Fig.~\ref{fig:BA Cartoon2}. $\alpha$1 is the unmodified control device currently used in BA tiles as seen on the right of Fig.~\ref{fig:BA Cartoon2}, which demonstrates the top view configuration of a current unreleased detector stackup. The nine $\alpha$ flavors investigate three independent mechanisms for softening $\alpha$ as described below. 

\begin{figure}[!t]
\centering
\includegraphics[width=0.45\linewidth]{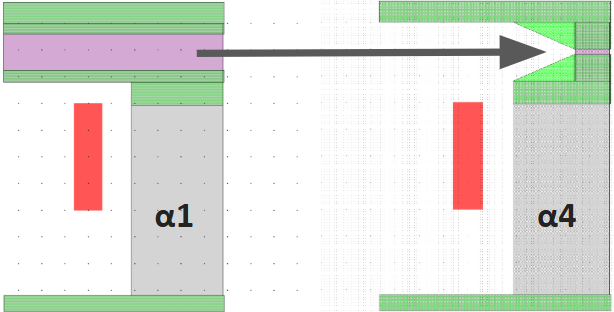}
\caption{$\alpha$1 and $\alpha$4 stackup top view comparison with the Al-plug (green), Ti TES (purple), and Al calibration TES (gray). Credit: JPL, A. Turner}
\label{fig:Alpha1 and Alpha4}
\end{figure}

$\alpha$2 to $\alpha$4 manipulate the TES size. They respectively vary the size of the TES area to $\frac{1}{2}$, $\frac{1}{4}$, and $\frac{1}{8}$ of the original design. The rest of the stack remains unchanged. The two dominant contributions that set the resistance of the TES are the uniform Ti film, and the Al-Ti interface. Devices where the Ti TES area is large cause the transition to be sharp, since the uniform Ti dominates the change in resistance. By decreasing the Ti area, the uniform Ti contribution lessens while the fixed contact contribution makes up an increasing share of the total change in resistance. The transition becomes shaped more by the broader contact behavior and less by the sharp uniform Ti. Fig.~\ref{fig:Alpha1 and Alpha4} shows $\alpha$1 next to $\alpha$4 to highlight the difference between the control detector and the most extreme Ti size reduction. 

$\alpha$5 and $\alpha$6 target a lower $\alpha$ by narrowing and lengthening the Al-Ti contact pathway, whilst maintaining the Ti TES geometry the same as the control device. 
A higher loop gain causes a smaller separation between the thermal and electrical time constants of the detector. The ratio, $\gamma$, between these two quantities is defined as:
\begin{equation}
\label{eq:gamma}
\gamma = \frac{G_{\text{int}}}{G_0},
\end{equation}
where $G_{\text{int}}$ is the thermal conductance between the Ti TES and the auxiliary heat capacity, and $G_{\text{0}}$ is the thermal conductance to the bath.  Their ratio (Eq.~\ref{eq:gamma}) is relevant as loop gain increases. 
When this separation becomes too small, the detector becomes underdamped and unstable, and consequently develops electrothermal oscillations \cite{IrwinHilton2005}.

\begin{figure}[!t]
\centering
\includegraphics[width=0.6\linewidth]{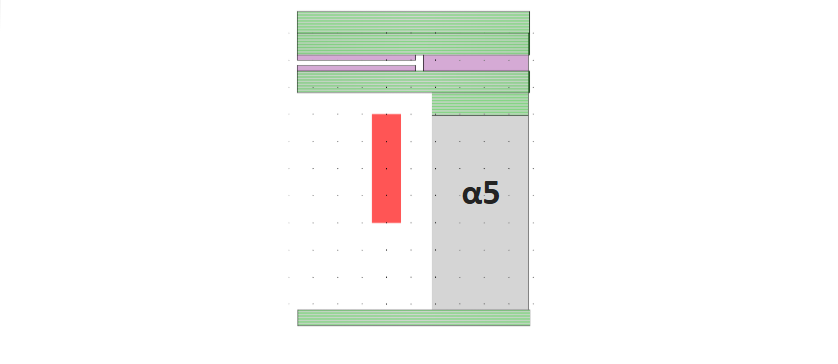}
\caption{$\alpha$5 stackup top view, narrowing and lengthening the Al-Ti contact pathway, whilst maintaining the Ti TES geometry the same as the control device. Credit: JPL, A. Turner.}
\label{fig:Alpha5}
\end{figure}

In $\alpha$5 and $\alpha$6, the Ti geometry is preserved, keeping $G_{\text{int}}$ high, whilst reducing the contact gap by a factor of around two. The two flavors differ only in the size of the reduction. Fig.~\ref{fig:Alpha5} shows the top view of $\alpha$5.

\begin{figure}[!t]
\centering
\includegraphics[width=0.5\linewidth]{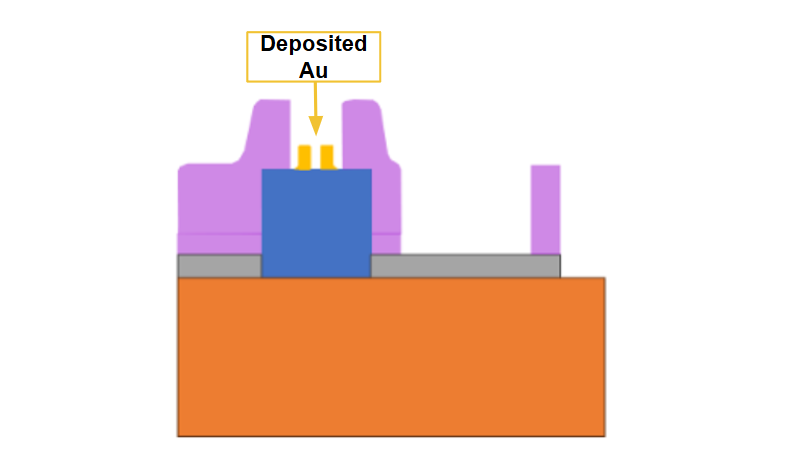}
\caption{$\alpha$7 through $\alpha$9 cross section cartoon where various patterns of normal metal Au at a specified thickness are deposited directly on top of the Ti TES. Credit: JPL, A. Turner}
\label{fig:Alpha7 to Alpha9 cartoon}
\end{figure}

\begin{figure}[!t]
\centering
\includegraphics[width=0.6\linewidth]{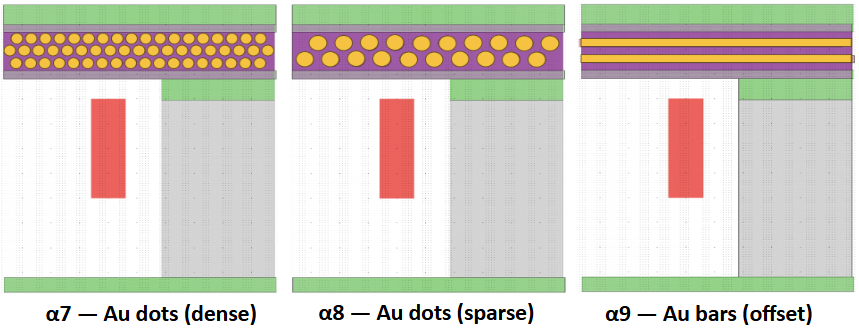}
\caption{$\alpha$7 through $\alpha$9 stackup top view. Credit: JPL, A. Turner}
\label{fig:Alpha7 to Alpha9 top view}
\end{figure}

$\alpha$7 through $\alpha$9 exploit the superconducting proximity effect with Gold (Au) normal metal. Various patterns of Au at a specified thickness are deposited directly on top of the Ti TES, as demonstrated in the cartoon model in Fig.~\ref{fig:Alpha7 to Alpha9 cartoon}. The three flavors use patterns of densely spaced small dots, sparser spaced larger dots, and horizontal bars for $\alpha$7 to $\alpha$9 respectively. Fig.~\ref{fig:Alpha7 to Alpha9 top view} shows the pattern variations across these $\alpha$ flavors. 

The Au is in direct contact with the superconducting Ti in certain, pattern-dependent regions. This causes the $T_c$ to become spatially inhomogeneous. The regions that are covered with Au enter transition at a lower temperature, whilst the uncovered areas have a higher $T_c$. This directly broadens the resistance-temperature $R(T)$ transition, and softens $\alpha$.

\subsection {Non-Inverse Stack Al-First Recipe Flavors}
\begin{figure}[!t]
\centering
\includegraphics[width=0.5\linewidth]{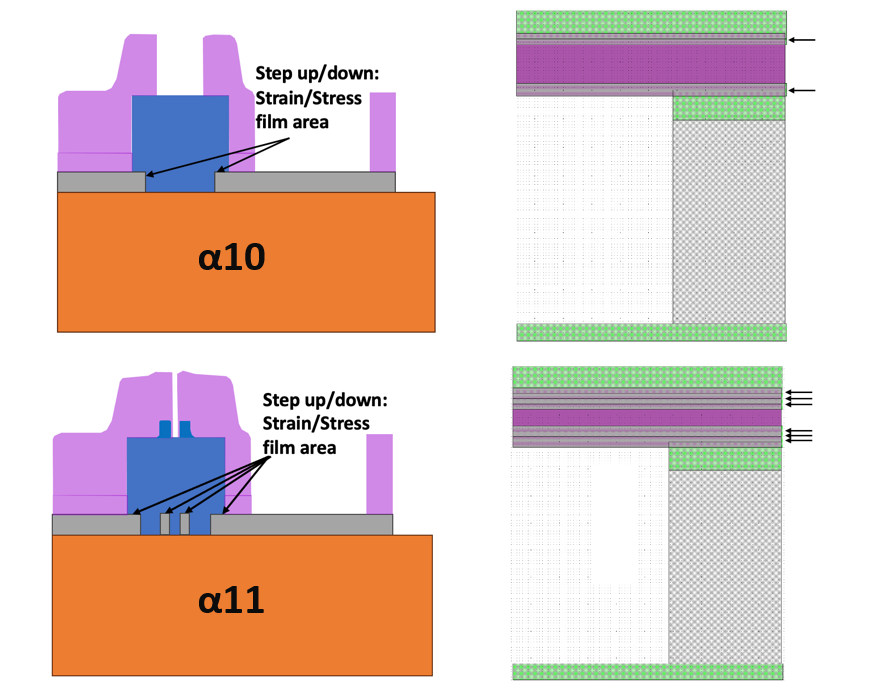}
\caption{$\alpha$10 and $\alpha$11 cartoon cross section (left) and stackup top view (right). $\alpha$10 and $\alpha$11 exploit the Al-film beneath the Ti TES to introduce height changes in the form of steps. Credit: JPL, A. Turner}
\label{fig:Alpha10 to Alpha11}
\end{figure}

All architectures in Wafer 2 ($\alpha$10 through $\alpha$16) use the Al-first recipe. This is the reversed fabrication order compared to Wafer 1 and enables architectural changes to be made beneath the Ti film. With this stackup, it is possible to exploit mechanical strain and stress effects and combine them with proximity effects. These architectures are investigated individually and combined. $\alpha$10 and $\alpha$11 exploit the Al-film beneath the Ti TES to introduce height changes in the form of steps. $\alpha$10 has two steps, and $\alpha$11 has six steps as seen in Fig.~\ref{fig:Alpha10 to Alpha11}.
The mask across both $\alpha$ types is the same, and they only differ in step count. As the Ti is deposited atop the raised Al-film areas, there are local areas of stress at the step edges. This strain in the Ti modifies the $T_c$ at each strained location, and so there is a population of modified $T_c$’s that differs from the unstrained regions. This mechanical modification introduces an inhomogeneity in $T_c$, broadening $R(T)$, and lowering $\alpha$.

\begin{figure}[!t]
\centering
\includegraphics[width=0.3\linewidth]{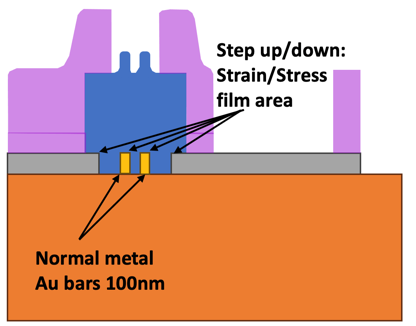}
\caption{$\alpha$12 cross section cartoon. Credit: JPL, A. Turner}
\label{fig:Alpha12 top view}
\end{figure}

$\alpha$12 and $\alpha$13 are similar to $\alpha$7 through $\alpha$9 from Wafer 1 in the way that they exploit the proximity effect with Au patterns. The difference is that these flavors are deposited to the opposite side of the film. $\alpha$12 (Fig.~\ref{fig:Alpha12 top view}) and $\alpha$13 vary from each other only by step count. $\alpha$12 deposits two Au bars and two Al-TES steps, and $\alpha$13 uses the same two Au bars, but without the additional Al-TES steps. Deposition of normal metal beneath the Ti addresses decreasing $\alpha$ on two fronts. The bars below the Ti introduce a physical step like $\alpha$10 and $\alpha$11 where the Ti is strained and forced to conform to the mechanical deviation. It additionally exploits the proximity effect since the Au bars are in direct contact with the Ti that sits on top of them. 

\begin{figure}[!t]
\centering
\includegraphics[width=0.85\linewidth]{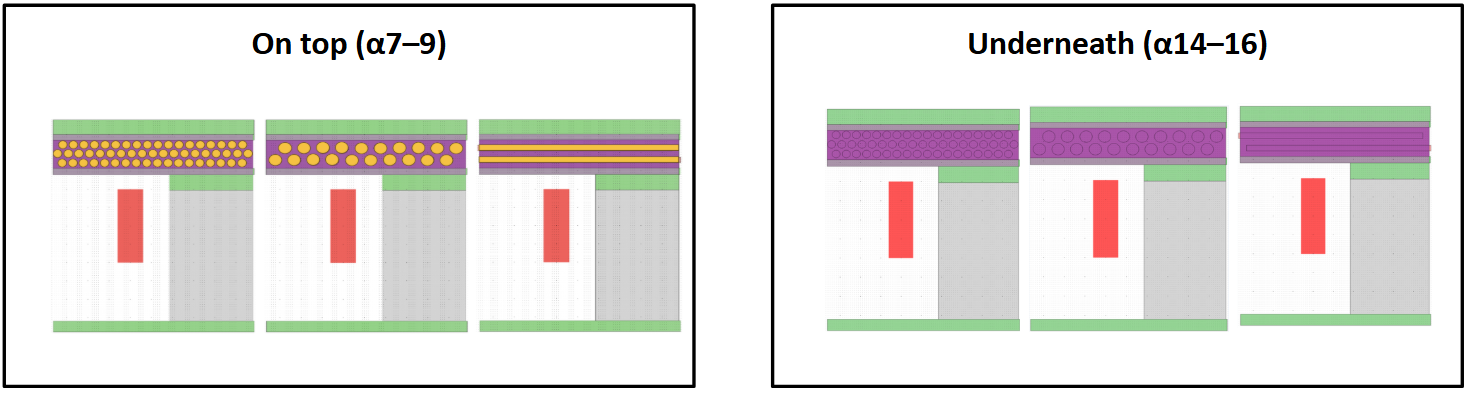}
\caption{$\alpha$14 to $\alpha$16 stackup (right) top view showing various patterns of Au deposited directly underneath the Ti TES compared against $\alpha$7 to $\alpha$9 stackup (left) where the Au is deposited atop the Ti TES. Credit: JPL, A. Turner}
\label{fig:Alpha14 to Alpha16 top view}
\end{figure}

As demonstrated in Fig.~\ref{fig:Alpha14 to Alpha16 top view}, $\alpha$14 to $\alpha$16 replicate the same Au patterns of $\alpha$7 through $\alpha$9 except that the Au is deposited beneath the Ti film. These TES flavors isolate whether the stack order itself affects the strength of the proximity-driven $\alpha$ reduction. 

\section{Methodology}
The wafers were tested at Caltech where continuous bias ramps were taken with temperature increasing and decreasing. For each flavor, $\alpha$ was extracted from fitting a sigmoid to the $R(T)$ transition.

The following results use the temperature down dataset as the primary reference. The two temperature sweep directions agree to within 17\%. The larger discrepancy is understood and exists because at the start of the sweep, the detectors are in different states. During the temperature increase, the detectors start in their superconducting state, so more power is required for the detectors to transition out from that state. However, when stepping down in temperature, the detectors are already in their normal state, so less power is required to get it back to the superconducting state.

\section{Results}
\begin{figure}[H]
\centering
\includegraphics[width=0.73\linewidth]{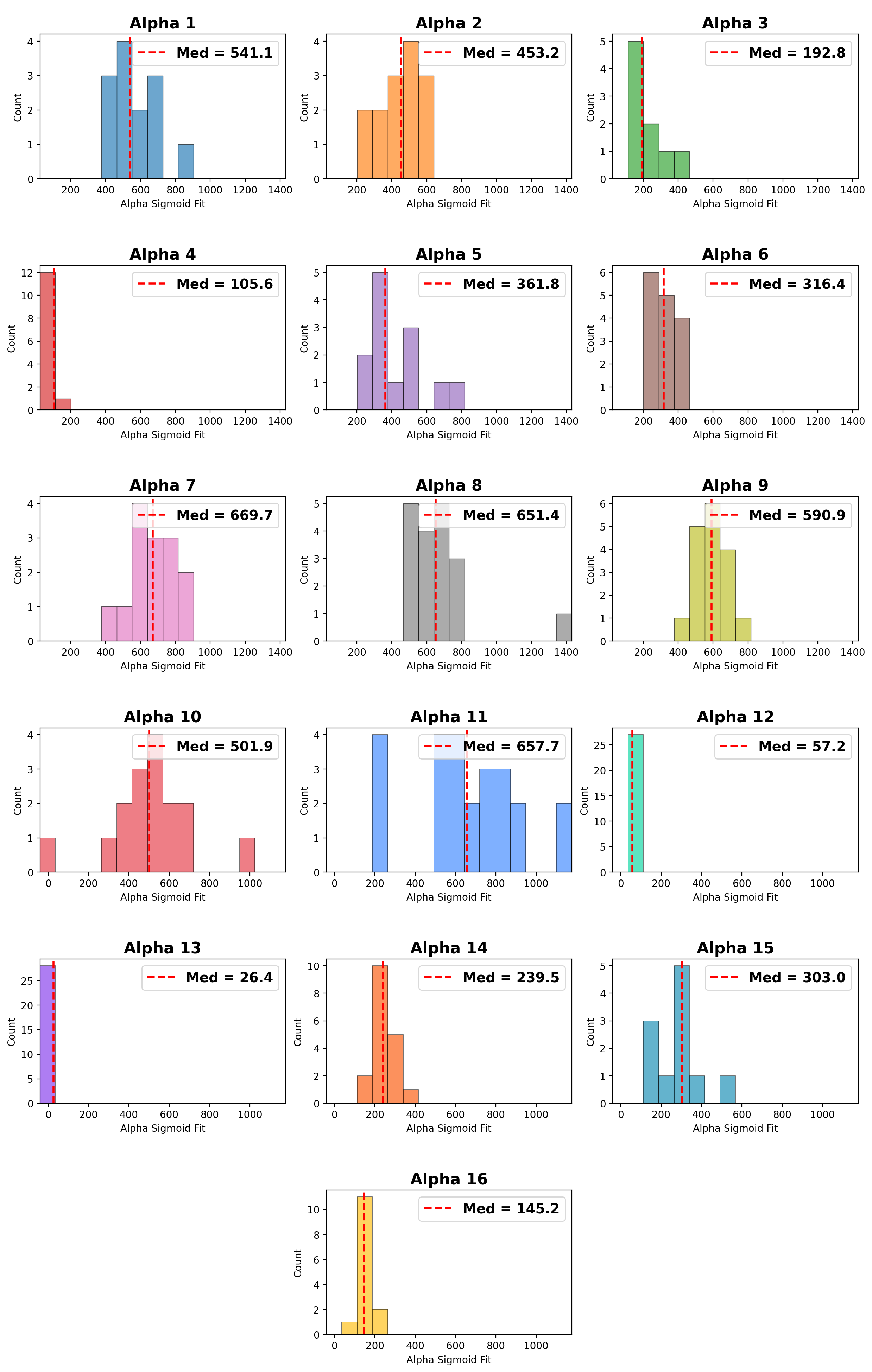}
\caption{Distribution of $\alpha$ results, where $\alpha$1 is the control and $\alpha$2 through $\alpha$16 are the various flavors. The vertical red dashed line indicates the median $\alpha$ value, which is also labeled in the top right corner for each flavor.}
\label{fig:alpha_histograms_combined}
\end{figure}

Fig.~\ref{fig:alpha_histograms_combined} shows the $\alpha$ distributions extracted from the sigmoid fits for all 16 architectures. Table~\ref{tab:alpha_summary} shows the median alpha value as well as the number of detectors tested. Across 16 flavors, $\alpha$ ranges from $\sim$26 to $\sim$670, confirming that the fabricated designs spanned a wide range in transition steepness. $\alpha$12 and $\alpha$13 show systematically the lowest $\alpha$ values ($\sim$57 and $\sim$26 respectively) of any grouping, which suggests that the dual attack of proximity effect and strain with the Al-first stack order proved most effective at reducing $\alpha$. While a lower $\alpha$ was desired, a ‘too low’ $\alpha$ would also not be viable since the bandwidth over which we could bias would be too small. The flavor with an $\alpha$ closest to the prescribed target value of 100 is $\alpha$4 at 105.6. The aggressively shrunk TES size ($\frac{1}{8}$) may pose fabrication and yield constraints. Because of this, $\alpha$3, ($\frac{1}{4}$ TES size) with an $\alpha$ of $\sim$193, is also a desirable option. An $\alpha$ $\sim$193 offers a robust factor of three reduction from the $\alpha$1 control and would not pose the same fabrication difficulties as $\alpha$4. $T_c$ was measured across all flavors and ranged 504–515 mK. The narrow band indicates that the softening mechanisms did not vary the bulk transition temperature significantly. 

\begin{table}[H]
\caption{Median $\alpha$ results to 3 significant figures, where $\alpha$1 is the control and $\alpha$2 through $\alpha$16 are the various flavors, and `n' is the number of detectors tested per flavor.}
\label{tab:alpha_summary}
\begin{tabular}{lcccccccccccccccc}
\toprule
Alpha & 1 & 2 & 3 & 4 & 5 & 6 & 7 & 8 & 9 & 10 & 11 & 12 & 13 & 14 & 15 & 16 \\
\midrule
N & 13 & 14 & 9 & 13 & 13 & 15 & 14 & 18 & 17 & 16 & 24 & 27 & 28 & 18 & 11 & 14 \\
Median & 541 & 453 & 193 & 106 & 362 & 316 & 670 & 651 & 591 & 502 & 658 & 57.2 & 26.4 & 239 & 303 & 145 \\
\bottomrule
\end{tabular}
\end{table}

\section{Future Work}
The $\alpha$ flavors fabricated in this test demonstrate that it is possible to tune $\alpha$ over more than an order of magnitude using TES fabrication recipe modifications. Independently from decreasing $\alpha$, we noted earlier that loop gain can be reduced by operating the focal plane at an elevated bath temperature in parallel. This method lowers the $P_{\text{electrical}}$ required to push the TES into transition, and further decreases the loop gain. This independent contribution to reduce loop gain means that a more modest reduction in $\alpha$ may be sufficient to meet noise requirements. The previously defined strict factor of five reduction may be loosened, and $\alpha$3 could satisfy requirements whilst yielding a more manufacturable and reproducible TES compared to $\alpha$4. Full detector arrays of both $\alpha$3 and $\alpha$4 released detectors are currently being fabricated. Results from these arrays will inform the TES architecture selected for the remaining tiles in BA3-220/270 and for the commissioning of the wafers for the BA4-90/150 receiver. 

\acknowledgments 
The BICEP/Keck experiments have been funded through U.S. National Science Foundation grants most recently including 2220444-2220448, 2216223, 1836010, and 1726917. The research was carried out at the Jet Propulsion Laboratory, California Institute of Technology, under a contract with the National Aeronautics and
Space Administration (80NM0018D0004). Focal plane development and testing were supported by the Gordon and Betty Moore Foundation at the California Institute of Technology. Readout electronics were supported by the Canada Foundation for Innovation grant to the University of British Columbia. The computations in this
paper were run on the Cannon cluster supported by the FAS Science Division Research Computing Group at Harvard University. The analysis effort at Stanford University and the SLAC National Accelerator Laboratory was partially supported by the Department of Energy. We thank the staff of the U.S. Antarctic Program and in
particular the South Pole Station without whose help this research would not have been possible. We also thank our winter-over operators: Manwei Chan, Karsten Look, Calvin Tsai, Paula Crock, Ta Lee Shue, Grantland Hall, Hans Boenish, Robert Schwarz, Sam Harrison, Anthony DeCicco, Thomas Leps, Brandon Amat, Nathan Precup,
Steffen Richter, Thibault Romand, Danielle Simmons, Markus Ayasse, Steven Jungst, Nathan McReynolds, and John Della Costa.

\end{document}